# Synthesis of a mesoscale ordered 2D π-conjugated polymer with semiconducting properties

G. Galeotti,[1,2,†,#] F. De Marchi,[1,#] E. Hamzehpoor,[3] O. MacLean,[1] M. Rajeswara Rao,[3] Y. Chen,[3] L. V. Besteiro,[1,4] D. Dettmann,[1,2] L. Ferrari,[2] F. Frezza,[2,5] P. M. Sheverdyaeva,[6] R. Liu,[7] A. K. Kundu,[6] P. Moras,[6] M. Ebrahimi,[1,‡] M. C. Gallagher,[7,*] F. Rosei,[1,*] D. F. Perepichka,[3,*] G. Contini[2,5,*]

1. Centre Energie, Matériaux et Télécommunications, Institut National de la Recherche Scientifique, 1650 Boulevard Lionel-Boulet, Varennes, Québec, Canada J3X 1S2
2. Istituto di Struttura della Materia, CNR, Via Fosso del Cavaliere 100, 00133 Roma Italy
3. Department of Chemistry, McGill University, 801 Sherbrooke Street West, Montreal, Québec, Canada H3A 0B8
4. Institute of Fundamental and Frontier Sciences, University of Electronic Science and Technology of China, Chengdu 610054, China
5. Department of Physics, University of Tor Vergata, Via della ricerca scientifica 1, 00133 Rome, Italy
6. Istituto di Struttura della Materia, CNR, S.S. 14, km 163.5, 34149 Trieste, Italy
7. Department of Physics, Lakehead University, 955 Oliver Rd, Thunder Bay, Ontario, Canada P7B 5E1

# These authors contributed equally to this work.

Current address: †Deutsches Museum, Museumsinsel 1, 80538 München, Germany, ‡Department of Chemistry, Lakehead University, 955 Oliver Road Thunder Bay, Ontario, Canada P7B 5E1.

* Corresponding author e-mail: mcgallag@lakeheadu.ca; dmitrii.perepichka@mcgill.ca; giorgio.contini@ism.cnr.it; rosei@emt.inrs.ca.

# Abstract

2D materials with high charge carrier mobility and tunable electronic band gaps have attracted intense research effort for their potential use as active components in nanoelectronics. 2D π-conjugated polymers (2DCP) constitute a promising sub-class due to the fact that the electronic band structure can be manipulated by varying the molecular building blocks, while at the same time preserving the key features of 2D materials such as Dirac cones and high charge mobility. The major challenge for their use in technological applications is to fabricate mesoscale ordered 2DCP networks since current synthetic routes yield only small domains with a high density of defects. Here we demonstrate the synthesis of a mesoscale ordered 2DCP with semiconducting properties and Dirac cone structures via Ullmann coupling on Au(111). This material has been obtained by combining rigid azatriangulene precursors and a hot dosing approach which favours molecular diffusion and reduces the formation of voids in the network. These results open opportunities for the synthesis of 2DCP Dirac cone materials and their integration into devices.

## Introduction

Since the isolation of graphene and the demonstration of its remarkable charge transport properties in 2004,[1] two-dimensional (2D) materials have become a major trending topic in materials science and nanotechnology.[2,3] Efforts have been devoted to identifying 2D materials beyond graphene that offer a greater degree of tunability and have a non-zero electronic band gap while retaining high carrier mobility. Among 2D materials, some present the rare property of Dirac cones in their electronic band structures, which give rise to ultrahigh carrier mobility that open new frontiers for fundamental research as well as technological development.[4] 2D π-conjugated polymers (2DCPs) can exhibit Dirac cones structures, with the advantage of customizable properties via structural modification (*i.e.* appropriate monomer design).[5-7] Heterotriangulene molecules are particularly interesting precursors for 2DCPs since Dirac semimetals or semiconductor materials can be obtained by shifting the Dirac cone position with respect to the Fermi level ($E_F$) through the choice of central atom or bridging functional groups, without perturbing other structural parameters (*i.e.* topology or unit cell dimensions).[5,8]

Dirac cones in the electronic band structure of 2DCPs have been predicted theoretically,[5,8] however experimental confirmation is still lacking.[9,10] The limited extent and high defect density of the 2DCP layers obtained so far has prevented the use of averaging techniques to measure their electronic band structure. The common approach to synthesize 2DCPs is surface-confined polymerization, where the surface acts both as a template and a catalyst for the coupling of the molecular precursors.[11-13] Ullmann coupling is the most common and general approach to on-surface synthesis of one-[14-18] and two-[19-24] dimensional polymers on metallic single crystal surfaces. Though it has been successfully applied in the formation of long 1D structures such as graphene nanoribbons (GNRs),[25,26] the largest domain sizes for 2DCPs synthesized to date have been less than 20×20 nm² due to defects, incomplete reactions, or polymorphism.[9,10,27]

Here, we present the synthesis of a mesoscale ordered 2DCP, with domain dimensions in excess of 100×100 nm² by Ullmann coupling of rigid, achiral heterotriangulene precursors deposited onto a preheated Au(111) surface. We used D3h symmetric tribromotriox*a*azatriangulene (TBTANG, Figure 1a) and tribromotriox*o*azatriangulene (TBTANGO, Figure 3f), which are predicted to form 2DCPs with semiconducting properties, and electronic band gaps of 1.8 and 2.4 eV respectively.[8] Both precursors are derivatives of triphenylamine (TPA), a well-known hole-transport material[28,29] used in optoelectronic devices.[30] The oxygen (for TBTANG backbone, TANG) or carbonyl (for TBTANGO backbone, TANGO) bridges ensure a planar geometry, improve π-conjugation (by precluding the dihedral twist of the phenyls rings) and modulate the electronic properties via electron-donating or withdrawing effects, respectively.[31] The P²TANG and P²TANGO polymers, obtained by polymerization of TBTANG and TBTANGO respectively, were characterized using scanning tunneling microscopy (STM), low-energy electron diffraction (LEED), and X-ray photoelectron spectroscopy (XPS) techniques, combined with statistical analysis to quantify the density of defects in the networks. The electronic band structure of the P²TANG was studied using angle-resolved photoelectron spectroscopy (ARPES) from a synchrotron source, revealing Dirac cones at the $\overline{K}$ point in the valence band, thus confirming the earlier density functional theory (DFT) predictions.[5,8] All these results demonstrate the unprecedented long-range order achieved for a 2DCP and provide guidelines to the synthesis of other mesoscale-ordered 2D polymers.

## Results

Dosing TBTANG onto a Au(111) surface kept above the dehalogenation temperature (180 °C) yields a highly ordered monolayer of large honeycomb networks which extend over the whole surface (Figure 1b,c). The pore-to-pore distance measured from the STM images is 1.73 ± 0.05 nm (Figure 1b), in close agreement with the calculated dimensions for P$^2$TANG (1.71 nm, Figure 1d). The honeycomb networks exhibit a low defect density, mainly single TBTANG vacancies (a statistical analysis of the defects is discussed later in the text). The domain dimensions are limited by the surface terrace size since step edges create a discontinuity in the network (Figure S1). The long-range ordered nature of the domains is confirmed by LEED patterns which show diffraction spots due to the polymer, rotated with respect to those of the substrate (Figure 1e-g). The P$^2$TANG diffraction spots closer to those of the substrate appear narrower and brighter (Figure 1e,f), while those further away are split in two distinguishable spots, showing the presence of two rotational domains of P$^2$TANG, with matrices $A_1 = \begin{pmatrix} 7 & 4 \\ -4 & 3 \end{pmatrix}$ and $A_2 = \begin{pmatrix} 7 & 3 \\ -3 & 4 \end{pmatrix}$, rotated by 25.29° and 34.71° with respect to the substrate lattice (Figure 1g and S2). These matrices give a unit cell parameter of 1.75 nm, in agreement with both STM (1.73 ± 0.05 nm) and DFT (1.71 nm) values. The two domains identified by LEED merge with a low number of defects, as shown by STM imaging (Figure S3), so that well-ordered continuous networks are observed across the entire substrate terraces.

The XPS analysis of P$^2$TANG/Au(111) shows four well-resolved components and a shake-up feature in the C 1s spectrum (Figure 1h). The four individual peaks are found at 285.7, 285.2 eV, 284.2 and 283.8 eV of binding energy (BE) and correspond to four inequivalent carbon atoms (C-O, C-N, C-C and C-H bonded) in the expected (2:1:1:2) ratio. Such remarkable resolution is achieved due to the extremely narrow widths of the peaks, a direct result of the high order of the polymer network. The presence of single components for the N 1*s* and O 1*s* spectra (Figure S4) and the intensity ratios of the components in the C 1s spectra are in agreement with the stoichiometry of P$^2$TANG, indicating that the TANG backbone remains intact following polymerization. Confirmation of complete polymerization is obtained by comparing the C 1s XPS spectra measured for P$^2$TANG and intact TBTANG, obtained on Au(111) at room temperature (RT), showing the disappearance of the peak at 284.8 eV of BE assigned to C-Br when C-C bonds are formed (Figure S5). The P$^2$TANG network reveals a remarkable thermal resistance and environmental stability, both of which are essential for practical device applications. No change in the STM images was observed after annealing P$^2$TANG/Au(111) at 400°C (Figure S6a). The hexagonal lattice remains intact after exposure to atmospheric conditions, as shown by STM and LEED (Figure S6b,c), with the expected presence of adsorbed atmospheric contaminants inside the honeycomb pores (Figures S6b). A comparison of the XPS spectra before and after atmospheric exposure reveals that these adsorbates are mostly removed by annealing the surface to 200°C in UHV (Figure S7). In addition, comparable order and atmospheric stability for P$^2$TANG is obtained when dosing TBTANG on expendable Au(111)/mica substrates (Figure S6d), which is essential for polymer transfer and subsequent device fabrication, as already demonstrated for GNRs.[25,26]

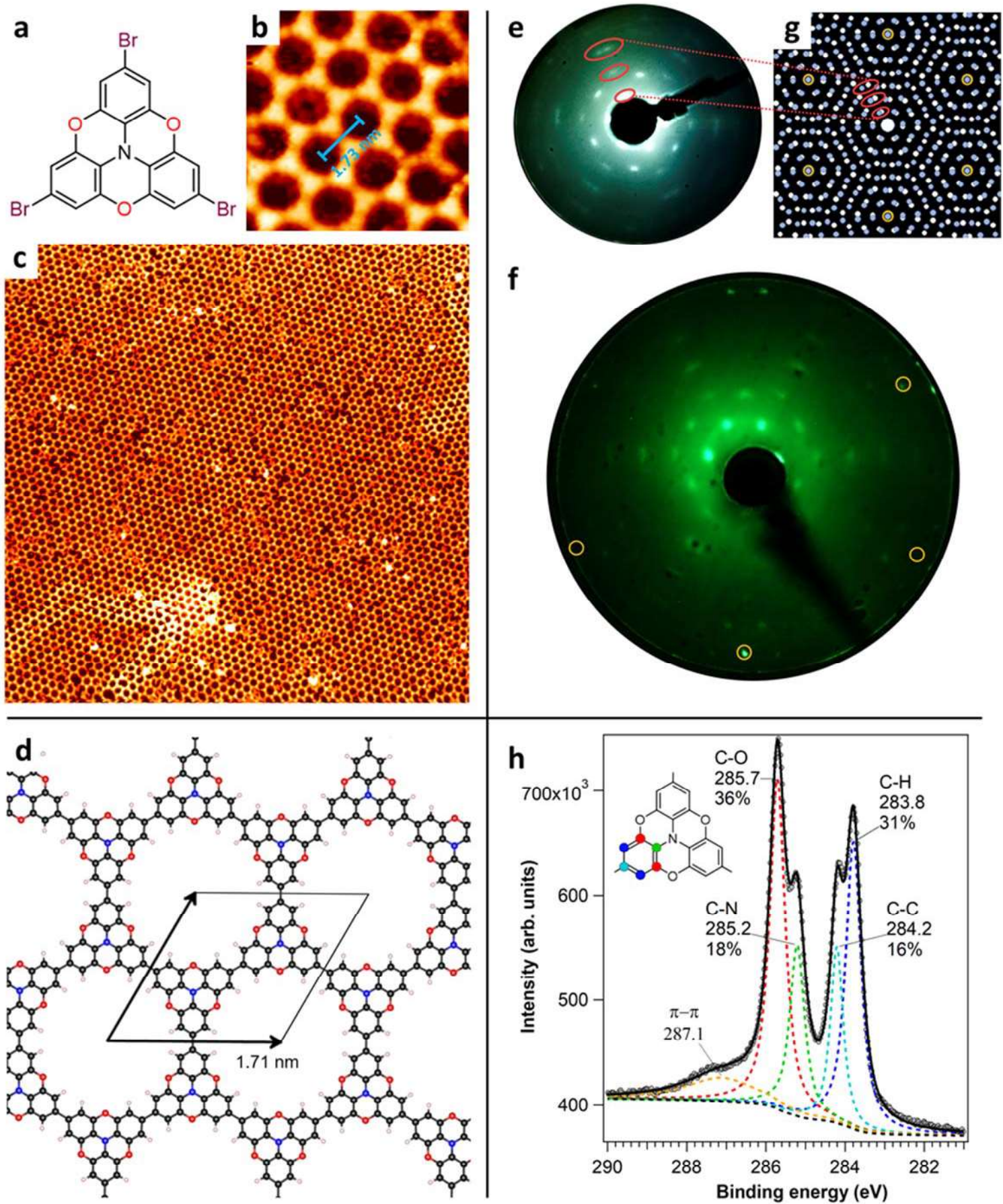

**Figure 1:** a) TBTANG molecular structure. b) 8x8 nm$^2$ and c) 90x90 nm$^2$ STM images of TBTANG dosed on Au(111) kept at 210 °C (It= -0.99nA, Vt= -1.12V). d) DFT simulated P$^2$TANG network, showing the unit cell. e-f) LEED pattern of P$^2$TANG/Au(111) at 210 °C, recorded at 42 eV (e) and at 72 eV (f); doubled spots are highlighted in e. g) Simulated LEED

pattern of the two superlattices $A_1$ and $A_2$. h) C 1$s$ XPS spectra (photon energy 450 eV) of P$^2$TANG/Au(111) at 210 °C, the color scheme of the TANG indicates the correspondence of C atoms and spectral components.

The mesoscale size of the P$^2$TANG network allows for the characterization of the electronic structure by ARPES. The dispersion of the valence band of the P$^2$TANG/Au(111) system has been investigated in the energy-momentum space along the $\overline{\Gamma KM}$ high symmetry direction of the surface Brillouin zone (SBZ) of Au(111), which corresponds to a mirror symmetry axis between the $A_1$ and $A_2$ domains of P$^2$TANG (Figure S8). The ARPES map of P$^2$TANG/Au(111) at 30 eV photon energy is presented in Figure 2a (ARPES of Au(111) is shown in Figure S9b). The calculated electronic band structure for a single domain P$^2$TANG shows bands with k-dispersion and the presence of Dirac cones at the $\overline{K}$-points above and below the $E_F$ (Figure S9e,g) and flat bands mainly localized on the O atoms[34] (Figure S9f,h). The high intensity of the 5$d$ Au bands hinders a detailed study of polymer related bands in the region between 2.5 and 8 eV of BE. P$^2$TANG-derived bands are visible outside this region at 2.00 eV, 10.00 eV, 12.28 eV, 14.50 eV and 16.44 eV of BE (black arrows in Figure 2a) in good agreement with the calculated electronic density of states (DOS) integrated along the momentum axis (Figure S9c,d).

The ARPES maps of P$^2$TANG/Au(111) and Au(111) in the proximity of $E_F$ are reported in Figure 2b-g. The $\overline{\Gamma KM}$ direction of Au(111) is not a high symmetry direction for the $A_1$ and $A_2$ rotational domains of the polymer. It coincides with a path that starts close to the ΓM direction of the P$^2$TANG SBZ (P-1 in Figure 2h) and crosses the polymer K point (P2) close to $\overline{K}_{Au}$, as shown in Figure 2h,i. Various features are visible in the ARPES maps, the most intense are due to the substrate bulk bands, namely the band crossing $E_F$ at about 1.06 Å$^{-1}$ and the one around $\overline{M}_{Au}$ (white arrows in Figure 2c), while the weaker features identified within 1 eV of $E_F$ are better visualized in the second derivative ARPES maps (Figure 2d,e). These structures are absent in the second derivative ARPES maps of Au(111) (Figure 2f,g) and, therefore, attributed to P$^2$TANG. Their wavelike dispersion does not depend on the photon energy, which is expected for 2D electronic states. The calculated band structure for freestanding P$^2$TANG (Figure 2j) is overlaid with the ARPES results in Figure 2d,e, showing an excellent agreement with the experimental data. The agreement of the observed band structure with the theoretical calculations further proves the formation of a long-range ordered semiconducting 2D polymer with extended π-conjugation.

A band with a cross-like shape, which resembles the Dirac cone predicted by DFT calculations (Figure 2j), is observed in the proximity of $\overline{K}_{Au}$, at the K (P2) point of the polymer (Figure 2d,e). A linear fit of this Dirac cone yields a band velocity of $v_F = 0.36·10^6$ m/s, which is slightly lower than the band velocity for graphene on the Au(111) surface ($v_F = 0.8·10^6$ m/s)[32] but higher than those calculated for a series of other hexagonal 2D polymers with C atoms as nodes and phenyls, thiophenes, pyrroles or pyrenes as linkers ($v_F < 0.15·10^6$ m/s).[5]

The interaction strength between P$^2$TANG and Au(111) can be gauged by examining the Shockley surface state of the Au(111). Such a state, which crosses the $E_F$ close to the $\overline{\Gamma}$ point, is known to be sensitive to the adlayer/substrate interaction and the variation of the energy minimum permits us to assess whether the adlayer is physi- or chemisorbed. The energy of the Shockley state band minimum is reduced from 0.47 eV[33] to 0.18 eV when P$^2$TANG is formed on Au(111) (Figure S10). This 0.29 eV energy shift is comparable to the 0.13-0.25 eV shift for P$^2$TANGO/Au(111) measured by STS[34] and somewhat higher than the 0.10 eV obtained for graphene/Au(111).[32] The effective mass, $m*/m_e$, of this Shockley state is 0.27, in line with values obtained for graphene/Au(111) and Au(111) (0.27 and 0.26, respectively).[32] The persistence of this surface state and its small shift in energy, together with an almost unchanged effective mass, suggests that the polymer is weakly bound and therefore the band structure measured by ARPES is not strongly influenced by the substrate, as indicated by DFT calculations (Figure S9).

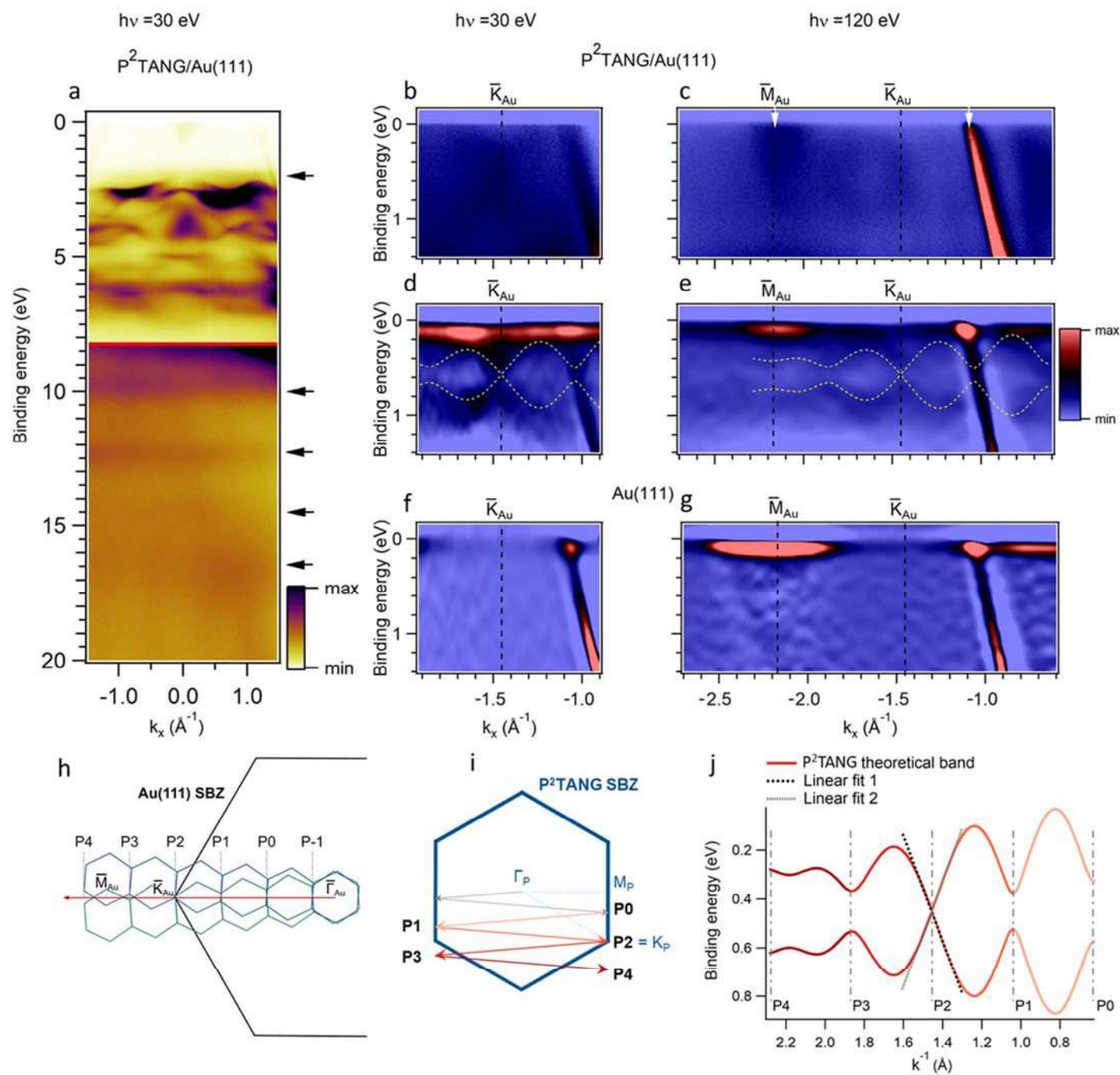

Figure 2: a-g) ARPES data of P$^2$TANG/Au(111) measured with photon energies of 30 eV (a,b,d,f) and 120 eV (c,e,g) at 50 K. a) The ARPES signal from the 8-20 eV region (below the red line) has been multiplied by a factor of 12 to highlight the experimental P$^2$TANG flat states, marked by black arrows. b-c) ARPES signal from P$^2$TANG/Au(111), and d-e) its second derivative map. The yellow dotted lines represent calculated bands for freestanding P$^2$TANG, shifted downward

by 0.12 eV to better match the experimental results. f-g) Au(111) second derivative ARPES map. h) SBZ scheme of the two $A_1$ and $A_2$ P$^2$TANG domains and Au(111) along the $\overline{\Gamma KM}$ direction. i) Path through the SBZ of one P$^2$TANG domain corresponding to the ARPES scan direction. j) Theoretical dispersion of freestanding P$^2$TANG HOMO states along the path shown in h,i.

## Discussion

The quality of the obtained P$^2$TANG network allowed for LEED and ARPES characterization for the first time of a 2DCP despite numerous related investigations.[9,10,13,27] The degree of order, the defect density and the morphology of the P$^2$TANG network have been characterized using the minimum spanning tree (MST) between the closed pore cells of the network.[35] Using this method we extract the mean edge distance (μ) and its standard deviation (σ) (Figure S11), the key values defining the network morphology.[36-38] Figure 3a compares the μ-σ pairs obtained for P$^2$TANG (green triangle) with other hexagonal 2D polymeric systems in the literature.[22,24,34,39-41] These results show that our networks are significantly more ordered than others reported previously, exhibiting a remarkably low standard deviation and μ-σ pair, which places them close to the perfect hexagonal lattice. The histogram of the nearest-neighbour distribution (Figure 3b) and a color-coded STM image (Figure 3a inset) both show that the majority of the pores (~80%) are hexagonal, while a much lower value (~40%) is obtained for the best data (indicated by red lines) currently available in the literature.[41]

The key to the improved network order is based on two aspects of the polymerization process: the high rate of surface diffusion relative to C-C coupling[22,42] and molecular precursor characteristics including rigidity, symmetry and achirality.

Most studies of surface polymer formation using Ullmann coupling follow a two step procedure, where the molecules are first deposited intact on a surface at RT and subsequently annealed to promote C-C bonding. Using this approach for TBTANG on Au(111) produces a high-density self-assembled lamellar structure at RT,[43] which consequently yields short disordered interdigitated oligomeric chains after annealing to 180 °C. This morphology is attributed to the initial surface density of precursors, which is too high to obtain the less-dense porous 2D P$^2$TANG network.

By dosing the precursor above the polymerization temperature, surface diffusion is enhanced, the high-density self-assembled phase is not formed, and a highly ordered 2D polymer is directly obtained. All experiments were performed using a low evaporation rate (about one hour to obtain a complete monolayer). This was selected to allow a lower probability of seed formation and the continuous growth of existing domains, as shown by Eichhorn et al.[24]

We find that the substrate temperature is pivotal, as the best result is obtained by keeping the Au(111) surface at 210 °C, while a 30 °C temperature variation noticeably reduces the quality of the network, as monitored by LEED (Figure 3c-e). Dosing at a lower Au(111) temperature (180 °C) yields the same LEED pattern with lower intensity diffraction spots. XPS analysis of samples prepared at 180 °C show that a C-Br contribution is still present, indicating that C-C coupling is limited by incomplete dehalogenation. At higher temperatures (240 °C), the LEED pattern shows the presence of an additional domain (Figure 3e and S13), aligned along the substrate and described by the epitaxial matrix $B = \begin{pmatrix} 6 & 0 \\ 0 & 6 \end{pmatrix}$. STM images (Figure S13e,f) show that non-hexagonal closed pores are mostly found at $A_{1,2} - B$ domain boundaries. The increased thermal energy facilitates the formation of such non-hexagonal pores with a higher enthalpy of formation, allowing for growth along multiple directions.

Importantly, hot dosing addresses the problem of hole formation. When the transport of the monomers to the growing polymer is limited to 2D diffusion, the formation of void defects in the network is unavoidable.[22] By hot dosing, incoming molecules are delivered uniformly across the surface and can adsorb and react directly in those voids, 'repairing' these defects.

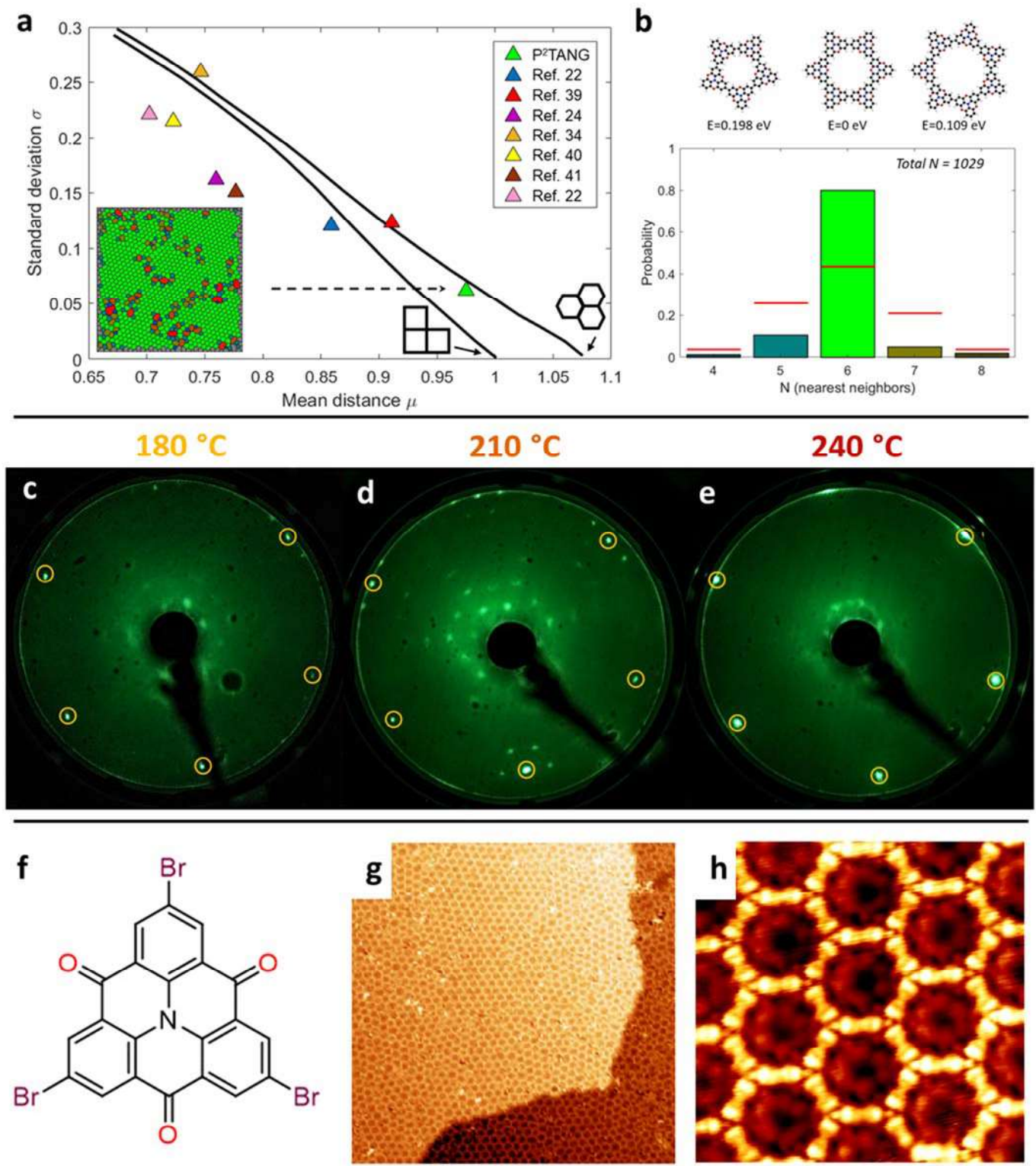

**Figure 3:** a) $\mu$ vs $\sigma$ of the MST's edge length. The data points were obtained from the STM images of different hexagonal porous networks; the two lines show the distortion trajectory from the ($\mu$, $\sigma$) point of perfect square and hexagonal lattices, at $\sigma = 0$, to a random distribution of pores. Our P²TANG lies on the hexagonal trajectory at lower $\sigma$ values compared with other work in the literature, quantitatively confirming the regularity of the polymer. The color-coded

STM image in the inset shows, in green, all pores in a perfect hexagonal lattice (*i.e.* with 6-neighbours). b) Bar diagram showing the fraction of polygons with i-sides. The calculated strain energy for the pentagon, hexagon and heptagon TANG oligomers is reported. We marked in red the bar diagram values for the best data points from the literature (Ref. 39 in a). c-e) LEED patterns of P$^2$TANG/Au(111) at various temperatures; the 210 °C experiment in d shows the best result, with long-range ordered polymers. f) TBTANGO molecular structure. g) 60 x 60 nm$^2$ large scale and h) 6 x 6 nm$^2$ detail of P$^2$TANGO dosed onto a hot surface at 210 °C (i: $I_t$=-0.16 nA, $V_t$=-0.26 V; j: $I_t$= 1.08 nA, $V_t$= 0.15 V).

The hot dosing approach was previously explored for other 2D polymeric systems, without yielding long-range ordered structures.[9,10,24,39] Most of this work used threefold molecules (triphenylbenzenes with Br and/or I leaving groups)[24,39,44] that have low strain energies for pentagonal and heptagonal defect formation (0.080 eV and 0.045 eV per monomer, respectively, as calculated by DFT), which are common in 2D polymers made with C$_3$ building blocks. In these studies, dosing at RT with subsequent annealing or dosing on a hot surface both led to defective networks, with about 40% of the molecules in an ideal hexagonal structure.[24,39] In our case, the success of the hot dosing approach can be (at least in part) attributed to the rigidity of TANG and TANGO nodes, which makes the formation of non-hexagonal structures unfavorable, as shown in Figure 3b and Figure S14 by the computed (high) strain energy per monomer for the pentagon (0.198 and 0.181 eV for TANG and TANGO, respectively) and heptagon (0.109 and 0.097 eV for TANG and TANGO, respectively). An increase of the strain energy can be obtained by adding functional groups to change the steric hindrance, or by using rigid monomers to increase the energetic cost of C-C bond bending. The first approach was explored by Fritton et al., who added ortho-methyl functional groups to increase the strain energy in non-hexagonal structures, obtaining highly ordered organometallic networks.[45] For TANG, the rigidity is due to the bridging oxygen, since non-bridged triphenylamine shows significantly lower computed strain energies for non-hexagonal rings (0.126 and 0.069 per monomer for the pentagon and heptagon, respectively, Figure S14).

The generality of our experimental approach for the realization of long-range ordered 2DCP with tunable electrical and mechanical properties was explored by depositing TBTANGO on Au(111) using the same surface temperature, leading to large area domains of ordered polymer (Figure 3g,h). These P$^2$TANGO domains are larger and possess a higher degree of order than those obtained by Steiner at al.[34] *via* a hierarchical approach using the same TANGO backbone with Br and I leaving groups.

## Conclusion and Perspectives

We have realized mesoscale ordered 2D π-conjugated polymers by dosing rigid achiral heterotriangulene molecules on hot (210 °C) Au(111) surface. This process minimizes the formation of network defects and favours the growth of highly extended (>100 nm) hexagonal structures. The substrate temperature is found to be a critical parameter as the highly ordered 2DCP could only be obtained within a 30 °C temperature range. The obtained polymer is resistant to annealing up to 400 °C and to air exposure.

The unprecedented order allowed for characterization of the 2D polymers using both local (STM) and surface averaging (XPS, LEED and ARPES) techniques, which were supported by DFT calculations and by statistical analysis to evaluate the defect density and morphology. ARPES measurements validate the previously theorized band structure and reveal a Dirac cone feature at the $\overline{K}$-point below $E_F$ and a band velocity of the same order of magnitude of that obtained for graphene on the Au(111) surface.

The mesoscale dimensions of the obtained 2DCP, their stability, the ability to grow on expendable Au(111)/mica substrates and the weak polymer/surface interaction, are all important for state-of-the-art device fabrication, thus opening a pathway towards the implementation of π-conjugated polymers with semiconducting properties for applications in electronics and photonics.

## Methods

The synthesis procedure for the TBTANG precursor follows that described by Suzuki et al.,[46] while the procedure for TBTANGO has been adapted from the one used by Fang et al.[47] (details in SI, section 10). All experiments were performed under ultrahigh vacuum (UHV) conditions with base pressures below $2\times10^{-10}$ mbar. Au(111) single crystals were cleaned by repeated cycles of Ar ion sputtering (1.0 keV for 10 minutes) and annealing (450 °C for 30 minutes). Both precursors were sublimed onto the Au(111) surface from a Knudsen cell (using a boron nitride crucible). During sublimation, the Au(111) sample was held at different temperatures, from RT up to 300 °C.

STM data were recorded with an Aarhus 150 STM by SPECS GmbH using etched tungsten tips. STM images were calibrated using the Au $\sqrt{3}\times22$ surface herringbone reconstruction by WSxM software.[48] Further image processing included plane subtraction and flattening.

XPS and ARPES measurements were performed at the VUV Photoemission beamline (Elettra Synchrotron, Trieste, Italy) on samples prepared *in situ*, using a Scienta R-4000 electron spectrometer at low temperature (50K) with p-polarized light. Between 30 eV and 120 eV photon energy, the energy and angular resolutions were set to 15 meV and 0.3°, respectively. XPS spectra were taken with 450 eV, 500 eV and 650 eV photon energies to collect C 1s, N 1s and O 1s, respectively, with total energy resolution of 30 meV.

XPS spectra were analyzed using Voigt lineshape peaks and Shirley backgrounds, while the BE scale was calibrated by using Au $4f_{7/2}$ at 84.00 eV and $E_F$.

Theoretical calculations were performed with the Vienna Ab-initio Simulation Package (VASP)[49] and Gaussian 16.[50] Plane-wave DFT calculations were made using the Pedrew-Burke-Ernzerhof[51] generalized-gradient approximation (PBE-GGA) for exchange-correlation potential, the projector augmented wave (PAW) method,[52] and a basis set with an energy cut off of 450 eV. The zero-damping DFT-D3 method of Grimme,[53] was used for van der Waals (vdW) correction of potential energy. A 3x3x1 k-mesh was used for geometry optimizations and a 9x9x1 k-mesh for electronic structure calculations. The ring strain energies were defined as the normalized gas-phase energies of the pentameric and heptameric rings relative to the hexamer, constrained in-plane to account for the effect of the substrate, and were computed with Gaussian 16 using the B3LYP functional and 6-31G(d) basis set.

The optimized Au lattice constant of 4.09820 Å was determined with a 12-atom Au(111) supercell and a 13x13x13 k-mesh. Calculations involving the Au substrate used a four-layer slab corresponding to the $\begin{pmatrix} 7 & 4 \\ -4 & 3 \end{pmatrix}$ epitaxy matrix, with a vacuum layer of 16 Å and the bottom two atomic layers fixed. The experimental Au(111) reconstruction was neglected on the basis of the weak polymer-surface interaction. Images of the simulated structures were generated using VMD software.[54]

## Author Contribution

G.G., F.D.M. and G.C. wrote the paper; G.G. and F.D.M performed laboratory XPS and STM experiments and analyzed the data; M.C.G., and D.D. participated in STM and laboratory XPS data acquisition and analysis; E.H., M.R.R. and Y.C. synthesized the molecules; O.M., E.H. and M.E. performed the DFT calculation; L.V.B. performed the statistical analysis of the STM images; G.G., G.C., P.M.S., P.M., A.K.K., F.F, and L.F. performed the synchrotron measurements, and G.G., G.C., P.M.S., P.M. and L.F. discussed and analyzed the data; G.G.,

D.D., F.F., R.L, M.C.G. and G.C. performed and analyzed the LEED experiments; G.C., M.C.G., F.R. and D.F.P. conceived the experiments and supervised the work; all authors participated in editing the manuscript.

## Supplementary Information

Supplementary information contains: XPS, ARPES and STM data and analysis, LEED images, DFT calculations.

## Conflict of interest

The authors declare no competing financial interest.

## Acknowledgments

This work was partially supported by a project Grande Rilevanza Italy-Quebec of the Italian Ministero degli Affari Esteri e della Cooperazione Internazionale (MAECI), Direzione Generale per la Promozione del Sistema Paese. M.C.G. D.F.P and F.R. acknowledge funding from the Natural Sciences and Engineering Research Council of Canada (NSERC) and the Fonds Québécois de la Recherche sur la Nature et les Technologies (FQNRT) through a Team Grant. Synthesis of the monomers was supported by US Army Research Office (grant W911NF-17-1-0126). F.R. is grateful to the Canada Research Program for funding and partial salary support. The authors acknowledge beamtime access and support from the Elettra light source in Italy. G.G., F.F. and G.C. thank N. Zema for laboratory support and useful discussions. Computations were performed mostly on the Niagara supercomputer at the SciNet HPC Consortium, funded by the Canada Foundation for Innovation under Compute Canada, the Government of Ontario, Ontario Research Fund – Research Excellence, and the University of Toronto, and in part on Graham and Cedar clusters of the Compute Canada. The simulations were also enabled by facilities of the Shared Hierarchical Academic Research Computing Network as well as WestGrid. A.K.K, P.M.S and P.M. acknowledge the project EUROFEL-ROADMAP ESFRI.